\documentstyle[aps,preprint]{revtex}
\begin{document}

\begin{center}
{\large Self-Organization of Value and Demand}
\vspace{1.0cm}

{\large R. Donangelo$^a$ and K. Sneppen$^b$}
\vspace{0.5cm}

$^a$ Instituto de F\'\i sica, Universidade Federal do Rio de Janeiro,\\
C.P. 68528, 21945-970 Rio de Janeiro, Brazil
\vspace{0.5cm}

$^b$ Nordita, Blegdamsvej 17, 2100 Copenhagen, Denmark

\end{center}
\vspace{0.5cm}

\begin{small}
\noindent Abstract: We study the dynamics of exchange value
in a system composed of many interacting agents.
The simple model we propose exhibits cooperative
emergence and collapse of global value for individual goods.
We demonstrate that the demand that drives the value exhibits
non-Gaussian ``fat tails'' and typical fluctuations
which grow with time interval as $\Delta t^H$, with $H\sim 0.7$.
\end{small}

\vspace{1.0cm}

The self organizational patterns of 
large non-equilibrium systems are of large current interest.
One aspect of such systems is their tendency to display
cooperative behavior, evidenced, for example, in the form of 
the occasional coordinated activity throughout the system 
\cite{BTW,SBFJ,Stange}.
Economic systems provide examples of such cooperative behavior,
that indeed show some of the characteristics of self organizing
systems. 
More specifically, agents reach a collective agreement about
what should be considered as valuable, and then use it as a
common trade object (money) \cite{Menger}.
Furthermore, the fluctuation of relative value with time
$t$ displays anomalous Hurst exponents 
(typical $\Delta D \propto \Delta t^H$ \cite{Hurst} with $H>1/2$) 
\cite{Mantegna,Evertz} and non-Gaussian statistics.
The latter fact was first noticed by Mandelbrot
\cite{Mandelbrot1,Mandelbrot2} and later quantified
in the observation that the succession of daily, weekly and
monthly distributions of exchange values possibly 
converges towards a Gaussian 
\cite{Akgiray,Ghashghaie}.

Recently a number of theoretical approaches have been developed to
deal either with the emergence of money as a cooperative phenomenon,
or with the non-Gaussian fluctuations associated to it.
In particular the work of Yasutomi \cite{Yasutomi} deals with the 
stochastic nature of emergence of certain goods as means of common
exchange between all the agents, {\it i.e.} how they become money.
Yasutomi makes the observation that if agents tends to accept what
others accept, then trade is facilitated. 
Therefore, emergence of money becomes related to history dependent
processes, which, as discussed by B. Arthur \cite{Arthur}
tend to lock the market into certain trade patterns.
This, however does not explain how one popular product may be replaced
by another, and thus result in an open and ever fluctuating system.
Yasutomi suggestion to solve this problem is to include both trade
costs and an evolving threshold for transactions.
In this way he obtains a bi-stable system where money emerges and
collapses quasiperiodically.
Dealing with the nature of fluctuations alone, Bak, Paczuski and
Shubik suggested a model with only one product, and
concluded that non-Gaussian fluctuations could be associated with the
crowding obtained when agents imitate each other's prize assessments
\cite{BPS}. They further obtained the anomalous Hurst exponents 
$H=0.65$ in a scenario
where local fluctuations are amplified by the global activity level.
Levy, Levy and Solomon \cite{LLS95} emphasize the effect of
heterogeneous expectations for the traders in a stock-bond market.
However, their specific model develops an unrealistic periodicity
for large systems \cite{Hellthaler}.
As we will see, our model self organizes heterogenity in a market
where each agent has its own limited history.
Furthermore, it predicts the emergence of non-periodic but 
persistent fluctuations in a market where many products compete
for attention.

In the present work we suggest a simple model that suggests that
emergence of ``money'' and its anomalous fluctuation in value 
are two sides of the same cooperative phenomena. 
We suggest that value emerges through
agents which make simple decisions based on 
their individual memory of earlier encounters
with other agents. The agents are not assumed to be smart,
the only trade that occurs is a one-good-for-one-good trade, 
and agents basically act in order to keep stock of all products.
The model leads to emergence and collapse of money,
it exhibits non-Gaussian statistics
and also displays long time fluctuations quantified by anomalous
Hurst exponents. 
\vspace{0.5cm}

The model we propose is as follows. 
We consider $N_{ag}$ agents and $N_{pr}$ different products.
Initially we give $N_{unit}$ units of the products to each agent.
The number $N_{unit}$ is fixed, but the products are chosen at
random, so the individuals are not in exactly the same situation.

At each timestep we select two random agents and let them attempt
to perform a trade between them.
The trade starts by comparing the list of goods that each agent
lacks and therefore would like to get from the other agent in
exchange for goods it has in stock.

We first consider the simple barter exchange procedure: when
each of the agents has products that the other needs, then one  
of these products, chosen at random, is exchanged.

In case such a barter exchange is not possible we consider
the ``money'' exchange procedure: one or both of the agents
accept goods which they do not lack, but consider useful for
future exchanges.

In order to determine the usefulness of a product, each agent $i$
keeps a record of the last requests for goods it received in
encounters with other agents. 
This memory is finite, having a length of $N_{mem}$ positions, 
each of which registers a product that was requested.
As the memory gets filled, the record of old transactions is lost.
Agents accept products they already have in stock with a probability
based on its memory record.
The chance of accepting such a good $j$ is taken to be proportional
to the number of times $T_{ij}$ that good $j$ appears on the memory
list of agent $i$:

\begin{equation}
p_{ij} \;=\; \frac{T_{ij}}{N_{mem}}\;,
\end{equation}
\vspace{0.5cm}

\noindent 
where we have used the fact that $\sum_j T_{ij} = N_{mem}$.
These two exchange mechanisms define our model.
\vspace{0.5cm}

In order to understand the dynamics of this model we first show, in
Fig.~1, the time evolution of the number of different kinds of trades.
The case depicted corresponds to $N_{ag}=200$ agents trading 
$N_{pr}=200$ types of products between them.
The memory list of each agent was chosen to be $N_{mem}=400$ items
long and the total ``richness'' of the economy was fixed by setting
the numbers of products possessed by each agent to $N_{unit}=400$.
This may be considered a typical set of parameters, and the behavior
is similar for a wide range of number of goods, memory sizes and
richness values, as long as memories are not too short and richness
not too high or too small.
If the economy is too rich, barter exchange disappears completely
as all agents will always have all types of goods.
On the other hand, if the economy is very poor, barter exchange
dominates completely and ``money'' exchange does not emerge.
Besides, in order for money to appear at all, the memory of each
agent must not be much shorter than the number of different products,
$N_{pr}$, (however, it can be much smaller than $N_{ag}$).

Fig.~1 shows that after a relatively short equilibration time the ratio
of barter to money transactions saturates.
The fact that money transactions dominate means that the agents have
already distributed their holdings in an efficient way, thus allowing
money exchange to become the dominating mode of transaction.
We also notice that the sum of the two types of exchanges
is less than 100\% , meaning that approximately 40\% of encounters
between agents did not lead to trades, for this set of parameters.

The interesting feature of our model is however not the relative number
of these two types of transactions, but the the fact that some products
become valuable as means of exchange.
In order to quantify this we define the monetary value of a good $j$ as
the number of agents $M=M(j)$ that consider this good to be the easiest
to trade.
In other words, each agent $i$ indicates the good $j$ with highest $p_{ij}$.

In Fig.~2a,b  we show the value of two particular goods as function
of time, which here we measure as number of encounters between agents.
Notice that the time scale for the evolution is a factor 100 larger
than that of Fig.~1.
This shows that, although the number of money and barter exchanges
has equilibrated almost instantaneously in the timescale of Fig.~2,
the evolution of the value of the goods displays an interesting
dynamics on this larger timescale.
In Fig.~2c we have plotted the number of agents that accept the most
desirable good.
From comparison with Figs.~2a and 2b, we notice that
the good in Fig.~2a becomes the most accepted in terms of exchange
early in the evolution of the system, and remains in that condition
until more than 900000 encounters have taken place, in this particular
history.
At this point, the good from Fig.~2b takes over, for a briefer
period, until time 1200000. 
There is a couple of additional crossovers between these two products,
but at the end another product arises and takes over as being the
most popular.

We stress that an important feature of the model is that often, and over
long periods, one particular good is considered valuable by a majority 
of the agents.
This appears without any {\it a priori} property of this product; the
value of the good develops through a cooperative agreement between the
agents about what is valuable.

How is this agreement reached?
In our model the agents only transfer information about what 
they need to replenish their stocks.
Thus the common concept of value emerges when many agents
need a good, {\it e.g.} because it is concentrated by some of the agents.
This accumulation is possible in our simple model because an agent accepts
goods even when having them in sufficient stock, if it feels they will be
useful for future transactions.
This property in itself makes a desired product to circulate, and possibly
concentrate, more than other goods, thereby making it more needed across
the system. 

However goods in high demand also collapse, as it was seen at time 
$\approx 800000$ in Fig.~2a.
This is because, due to the large number of money transactions, the most
valuable good may, through random fluctuations, become better distributed
than some other good, which then replaces it as the system's money.

We now try to quantify these fluctuations.
Fig.~3a displays the changes in the value of $M$, which time evolution
was seen in Fig.~2c, for various time intervals $\Delta t$.
The figure shows that, for sufficiently large values of $\Delta t$, these
fluctuations have exponential tails.
This is in contrast to the normal Gaussian behavior expected for
value assigned by independent agents.
The fluctuations obtained for the different values of $\Delta t$ could not
be made to collapse into one curve. 
This signals that the short and long time statistics 
of $M$ cannot be described by the same Hurst exponent.

In order to examine the statistics of the
underlying demand of goods we show, in Fig.3b, the fluctuations
in the number of times $D$ that a given product $j$ appears in the
memory of the system,  

\begin{equation}
D(j) \;=\; \sum_i T_{ij}\;,
\end{equation}

\noindent where the sum runs over all agents $i$ in the system.
In Fig.~3b we show a data collapse of fluctuations in $D$, {\it i.e.} 
$P( \Delta D ) \rightarrow \Delta t^H \cdot P( \frac{\Delta D}{\Delta t^H} )$
for 3 different values of $\Delta t$.
We notice that the curves collapse with Hurst exponent $H=0.70$,
an observation that we confirmed by finding numerically that 
$\left( \langle (D(t+\Delta t)-D(t))^2\rangle \right) ^{\frac {1}{2}} 
\propto \Delta t^{0.68 \pm 0.02}$, over more than three 
orders of magnitude. 
The fact that $H>0.5$ indicates that fluctuations 
in demand exhibit persistency,
{\it i.e.} that trends are amplified by a 
self organized cooperative feedback in our model. 
This relatively large anomalous Hurst exponent 
requires that there is a sizeable number of different products.
We have found that persistence exists for nearly all
system sizes.
The value of the Hurst exponent was found to 0.6 for a size 10 system
whereas it was found to be 0.7 for system size 100 and 1000.
For size dependence we mean here that we scale equally the parameters
of the model: $N_{ag}, N_{pr}, N_{mem}, N_{unit}$.
We have also checked that a change in $N_{ag}$ alone from 100 to 1000
also leads to a Hurst exponent of 0.7.
Only in the case of $N_{pr}=2$ we found $H=0.5$, and this for any value
of the other parameters.

We further notice, from Fig.~3b, that the fluctuations exhibit fat tails. 
In particular, the short time scale statistics tend to have exponential
tails whereas the long timescale fluctuations are more Gaussian-like.
The difference of Hurst scaling between value $M$ and demand $D$ reflects the
fact that changes in the most wanted product are faster, but less persistent
than structural changes in the composition of demand.

We have examined variations of the above model, as, for example, including
reluctance for the agents to trade away goods they consider valuable,
{\it i.e.}
with high $p_{ij}$.
Also we have considered the possibility that agents include in $T_i$ only
trades that were effectively performed, and not just requests. 
Finally, we have tested the case where the $T_i$ lists also include requests
for ``money trades'', arising from what other agents consider valuable.
All these cases change the information exchanged between agents about what is
valuable. However, the qualitative behavior of our model 
was in all cases similar, indicating its robustness. 

It is tempting to compare the emergence of value in the above model
with the emergence of monetary systems in the real world.
A beautiful historical example of the collapse of one currency and
the appearance of a new one in a XVII century chinese town is given
in Yasutomi's work \cite{Yasutomi}. 
More familiar to the reader may be similar events in the market for
tulips in the Netherlands, or the market for gold in more recent times. 
In all these cases the acceptability of the good or currency may drop
dramatically, without any deeper reason than the fact that nobody
considers them valuable any more.
More quantifiable are the values of currencies, where now, due to events
that have taken place in this century, the U.S. Dollar has become
globally accepted. 
In that regards one should mention that fluctuations in monetary value 
seem to exhibit fat tails \cite{Evertz,Mandelbrot1}, and long time
fluctuations that can be characterized by a Hurst exponent
$H\approx 0.55$ \cite{Evertz}. 
All these features are consistent with our simple scenario. 
We stress that our model is schematic and does not include any development
of strategy by the agents, strategies which would naturally influence the
emergence of cooperativity \cite{Zhang,Paczuski}. 

Furthermore, the present version of the model does
not include effects related to production or consumption of goods.
We have verified that the main result of including these processes,
is that products that are easily produced never become valuable.
Thus, if all goods are produced and consumed at a high rate,  no common
``money'' will emerge.
On the other hand, if products are produced at a low rate the results
presented above remain valid, indicating again the robustness of the
present model. 
In fact, the addition of a slow production and consumption of goods
allows the system to settle into a statistical stationary state.
This avoids products from falling, one by one, into the absorbing
state consisting of having the product present in the stock of all
agents, and, consequently, being gradually removed from the memories.
As soon as such a product is removed from all memories, it is
effectively not anymore present in the economy.
The effect of the decay mentioned above is negligible in the time
scale of the calculations presented in this work.

Finally we would like to emphasize that our work purports to model
behavior in a social setting where information exchange is a key feature. 
The fact that some forms of information exchange lead to self-organization
has already been proposed by {\it e.g.} Bonabeau {\it et al.} 
\cite{Bonabeau95}.
There sociological hierarchies emerge through comparisons of a dominance
index assigned to each individual.
Our model is more adapted for market behavior, in particular because it
also considers the exchange of goods, and not only of information.
By emphasizing the interplay between exchange of products and information,
we have constructed a  minimalistic market model.
However, since we explicitly require that our agents do not develop
different strategies, and that all the products can be exchanged only
in a one-to-one manner, our model may not give answers to some
questions that may be adressed in more detailed simulation models
{\it e.g.} employing minority games \cite{Zhang}, or the more 
elaborate models of strategy and investments programs also existing
in the economic literature, {\it e.g.} the one by Kim and Markowitz
\cite{KM89}. 

In summary, we have constructed a simple model for the 
cooperative concept of money. The model suggests that money 
emerges and collapses as a simple consequence of trade of 
goods between interacting agents with memory.
The concept of value arises from the probability that a local 
dynamical pattern (the need for certain products by individual 
agents) results in a global one (the general acceptability of 
products for exchange). An important consequence, possibly 
valid for the dynamics of other large systems, is that a 
tendency to draw information from previously encountered 
patterns may result in a  dynamic behavior that exhibits 
non-Gaussian fluctuations and persistency, evidenced by 
anomalous Hurst exponents.
\vspace{0.5cm}

We thank Maya Paczuski and Dietrich Stauffer for constructive
comments to the manuscript.
R.D. thanks Nordita for its hospitality and financial support during
the time that this work was performed.

\vspace{0.5cm}
{\bf Figure Captions}

\begin{itemize}

\item Fig. 1. Probability, measured as a percentage of the encounters
between agents that result in barter (full line) or money exchanges
(dashed line), as defined in the text.

\item Fig. 2. Number of agents $M$ that consider a particular product as
the most valuable (a and b), and the maximum number of agents that, at a
given time, consider the same product as the most valuable (c).

\item Fig. 3. (a) Probability of having changes in the value $M$ as a
function of their size, for the three different time steps
$\Delta t =200000$ (full line),
$\Delta t = 20000$ (long dashed line), and
$\Delta t = 2000$ (short dashed line).
(b) Probability of having changes in the demand $D$. The demand is here
measured in units of $\Delta D/\Delta t^H$, where $\Delta t$ takes the
values $200$, $2000$, and $20000$ (three thin lines), and we took for the
Hurst exponent the value $H = 0.7$.    
The thick line depicts the normal probability distribution having the
same mean and standard deviation as that for $\Delta t = 200$.
The peaks near 0 observed in the other distributions are attributed
to border effects, arising from transitions near $D=0$.
\end{itemize} 

\end{document}